\documentclass{vldb}
\usepackage{balance}  

\newcommand{\eat}[1]{}

\usepackage{latexsym}
\usepackage{amsfonts}
\usepackage{amsmath}
\usepackage{amssymb}
\usepackage{color}
\usepackage{colortbl}
\usepackage{epsfig}
\usepackage{xspace}
\usepackage{graphicx}
\usepackage{subfigure}
\usepackage{paralist}
\usepackage{enumerate}
\usepackage[color,matrix,arrow,all]{xy}
\usepackage{comment}
\usepackage{booktabs}
\usepackage{balance}
\usepackage{stmaryrd}
\usepackage{pifont}
\usepackage{hhline}
\usepackage{listings}
\usepackage{array}
\usepackage{float}
\usepackage[flushleft]{threeparttable}

\usepackage{xcolor}

\newcolumntype{L}[1]{>{\raggedright\let\newline\\\arraybackslash\hspace{0pt}}m{#1}}
\newcolumntype{C}[1]{>{\centering\let\newline\\\arraybackslash\hspace{0pt}}m{#1}}
\newcolumntype{R}[1]{>{\raggedleft\let\newline\\\arraybackslash\hspace{0pt}}m{#1}}

\usepackage{epsfig}
\usepackage{multirow}
\usepackage{url}

\usepackage[export]{adjustbox}
\usepackage[noend]{algorithmic}
\usepackage{algorithm}

\newcommand{\add}[1]{\textcolor{blue}{{#1}}}

\newcommand{\at}[1]{\protect\ensuremath{\mathsf{#1}}\xspace}

\newcommand{\sstab}{\rule{0pt}{8pt}\\[-2.2ex]}

\newcommand{\bi}{\begin{itemize}}
\newcommand{\ei}{\end{itemize}}

\newcommand{\be}{\begin{enumerate}}
\newcommand{\ee}{\end{enumerate}}
\newcommand{\beqn}{\begin{eqnarray*}}
\newcommand{\eeqn}{\end{eqnarray*}}

\newcommand{\stitle}[1]{\vspace{1ex}\noindent{\bf #1}}

\newcommand{\etitle}[1]{\vspace{1ex}\noindent{\underline{\em #1}}}

\newcommand{\ie}{{\em i.e.,}\xspace}
\newcommand{\eg}{{\em e.g.,}\xspace}

\newcommand{\aka}{\emph{a.k.a.}\xspace}

\newcommand{\False}{\mbox{\em false}}
\newcommand{\True}{\mbox{\em true}}

\newcounter{ccc}

\newcommand{\DRs}{{\small DRs}\xspace}
\newcommand{\DR}{{\small DR}\xspace}

\DeclareMathAlphabet{\pazocal}{OMS}{zplm}{m}{n}

\newcommand{\eop}{\hspace*{\fill}\mbox{$\Box$}\vspace{1ex}}     

\newcounter{example}
\renewcommand{\theexample}{\arabic{example}}
\newenvironment{example}{
        \vspace{1ex}
        \refstepcounter{example}
        {\noindent\bf Example \theexample:}}{
        \eop}

\newcommand{\nthesection}{\arabic{section}}
\newcounter{definition}[section]
\newenvironment{definition}{
        \vspace{1ex}
        \refstepcounter{definition}
        {\noindent\bf Definition {\bf \thedefinition}:}}{\eop
}

\newcounter{alg}[section]
\renewcommand{\thealg}{\nthesection.\arabic{alg}}

\newcounter{arule}
\renewcommand{\thearule}{\arabic{arule}}

\newcounter{claim}
\renewcommand{\theclaim}{\arabic{claim}}

\newcommand{\sys}{{\sc DeepER}\xspace}

\makeatletter
    \newcommand\figcaption{\def\@captype{figure}\caption}
    \newcommand\tabcaption{\def\@captype{table}\caption}
\makeatother

\definecolor{shadecolor}{RGB}{200,200,200}

\definecolor{shadecolor1}{RGB}{230,230,230}

\definecolor{shadecolor1}{RGB}{255, 114, 118}

\vldbTitle{Distributed Representations of Tuples for Entity Resolution}
\vldbAuthors{Muhammad Ebraheem, Saravanan Thirumuruganathan, Shafiq Joty, Mourad Ouzzani, and Nan Tang}
\vldbDOI{https://doi.org/10.14778/3236187.3236198}

\title{Distributed Representations of Tuples for Entity Resolution}

\author{
	\hspace{-3ex}
	Muhammad Ebraheem
	~Saravanan Thirumuruganathan
	~Shafiq Joty$^\dag$
   	~Mourad Ouzzani
	~Nan Tang
	\vspace{.5ex}
	\\
	\affaddr{Qatar Computing Research Institute, HBKU, Qatar \qquad $^\dag$Nanyang Technological University, Singapore}
	\\	
	\{mhasan, sthirumuruganathan, mouzzani, ntang\}@hbku.edu.qa, srjoty@ntu.edu.sg
}

\begin{document}

\pagestyle{empty}

\maketitle

\begin{abstract}
Despite the efforts in $70+$ years in all aspects of entity resolution (ER), there is still a high demand for democratizing ER --  by reducing the heavy human involvement in labeling data, 
performing feature engineering, tuning parameters, and defining blocking functions.
With the recent advances in deep learning, in particular distributed representations of words (\aka word embeddings), we present a novel ER system, called \sys, that achieves good accuracy, high efficiency, as well as ease-of-use (\ie much less human efforts).
We use sophisticated composition methods, namely uni- and bi-directional recurrent neural networks (RNNs) with long short term memory (LSTM) hidden units, to convert each tuple  to a distributed representation (\ie a vector), which can in turn be used to effectively capture similarities between tuples.
We consider both the case where pre-trained word embeddings are available 
as well the case where they are not; we present ways to learn and tune the distributed representations that are customized for a specific ER task under different scenarios.
We propose a locality sensitive hashing (LSH) based blocking approach that takes all attributes of a tuple into consideration and produces much smaller blocks, compared with traditional methods that consider only a few attributes.
%
%
We evaluate our algorithms on multiple datasets (including benchmarks, biomedical data, as well as multi-lingual data) and 
the extensive experimental results show that \sys outperforms existing solutions.
\end{abstract}

\section{Introduction}
\label{sec:introduction}

Entity resolution (ER) (\aka record linkage), a fundamental problem in data integration, has been extensively studied for $70+$ years~\cite{recordlinkage}, 
from different aspects such as 
declarative rules~\cite{DBLP:conf/kdd/BilenkoM03,synthesizer,DBLP:journals/pvldb/WangLYF11}, 
machine learning (or probabilistic)~\cite{theoryrecordlinkage,DBLP:conf/kdd/BilenkoM03,SarawagiB02,DBLP:conf/kdd/CohenR02}, and 
crowdsourcing~\cite{DBLP:journals/pvldb/WangKFF12,corleone}, to name a few,
and in many domains such as health care~\cite{recordlinkage}, e-commerce~\cite{corleone}, data warehouses~\cite{DBLP:reference/dataware/Winkler09}, and many more.

\begin{figure}[t!]
	\centering
	\includegraphics[width=\columnwidth]{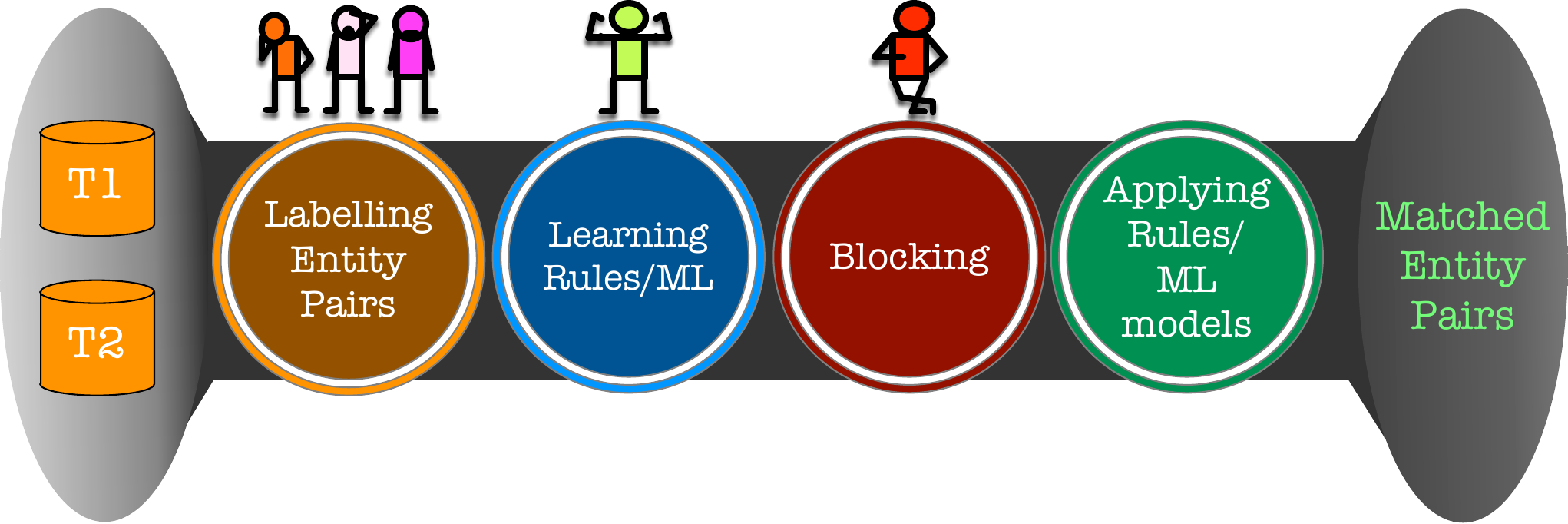}
	\vspace*{-5ex}
	\caption{A Typical ER Pipeline}
	\label{fig:er}
	\vspace*{-3ex}	
\end{figure}

Despite the great efforts in all aspects of ER, there is still a long journey ahead in democratizing ER. 
Adding to the difficulty is the rapidly increasing size, number, and variety of sources of big data.
Consider Figure~\ref{fig:er} for a typical ER pipeline that consists of four main steps:
(1) {\em labeling entity pairs} as either matching or non-matching pairs;
(2) {\em learning rules/ML models} using the labeled data\footnote{Rules can also be hand-crafted based on domain knowledge.}; 
(3) {\em blocking} for reducing the number of comparisons; and
(4) {\em applying} the learned rules/ML models.
Step~(1) decides {\em what} are the matched entities.
Step~(2) reasons about {\em why} they match. 
Step~(3) reduces the number of pairwise comparisons for step~(4)
(\ie~{\em how}), usually by expert specified blocking functions, 
which generate blocks such that matched entities co-exist in the same block.

\stitle{Challenges.}
The major challenge of current solutions in democratizing ER is that each step needs human-in-the-loop. 
Even a ``simple'' step, such as step~(1), which is thought to be trivial, turned out to be difficult in practice~\cite{DBLP:conf/sigmod/DoanABDGKLMPCZ17}.
Moreover, the human resources required in each step might be different -- knowing what (\ie step~(1)) is easier than telling why (\ie step~(2)) or how (\ie step~(3)).
Wouldn't it be great if we significantly reduce human cost for each step, but with a comparable if not better accuracy?

\stitle{Observations.}
(i) In practice, step~(1) is tedious because humans can only label up to several hundred (or few thousands) entity pairs and are error-prone. 
Intuitively, the hope to reduce this effort is to have a ``prior knowledge'' about what values would most likely match.
(ii) Regardless of using rule-based~\cite{synthesizer,DBLP:journals/pvldb/WangLYF11} or ML-based~\cite{theoryrecordlinkage,DBLP:conf/kdd/BilenkoM03} methods, 
step~(2) requires experts to provide (domain-specific) similarity functions 
from a large pool, for example, SimMetrics (\url{https://github.com/Simmetrics/simmetrics}) has $29$ symbolic similarity functions.
In addition, experts may also need to specify the thresholds.
Ideally, this step needs a unified metric that can decide different cases of matched entities, from both  syntactic and semantic perspectives.
(iii)~For step~(3), a blocking function is typically defined over few attributes, \eg~\at{country} and \at{gender} in a table about demographic information, 
without a holistic view over all attributes or the semantics of the entities.

\stitle{Our Methodology.}
We present \sys, a system for democratizing ER that 
needs much less labeled data by considering   prior knowledge of matched values (observation~(i)), 
captures both syntactic and semantic similarities without feature engineering and parameter tuning (observation~(ii)), 
and provides an automated and customizable blocking method that takes a holistic view of all attributes (observation~(iii)) -- all of these targets are achieved by gracefully using 
distributed representations, or \DRs for short, (of tuples), a fundamental concept in deep learning (DL). 

\DRs of tuples is an extension of \DRs of words (\aka word embeddings) where a word is mapped to a high dimensional dense vector such that the vectors for similar words are close to each other in their semantic space. 
Well known methods include word2vec~\cite{mikolov2013distributed} and GloVe~\cite{pennington2014glove}.
Word embeddings have become conventional wisdom in other fields such as NLP and have many appealing properties:
(a)~They are known to well capture semantic string similarities, \eg ``William'' and ``Bill'', and ``Apple Phone'' and ``iPhone''.
(b)~Using pre-trained word embeddings (\eg GloVe is trained on a corpus with 840 billion tokens), we can tremendously  reduce the human effort of labeling matched values per dataset.
(c)~Due to their generality, it is possible that the same intermediate representation (such as from word2vec, GloVe, or fastText)  can work for multiple datasets out-of-the-box.  
In contrast, traditional ER approaches require hand tuning for each dataset.
(d)~\DR toolkits such as fastText~\cite{bojanowski2016enriching} provide support for almost 294 languages allowing them, and thereby \sys, to work seamlessly on different languages.
(e)~They provide a new opportunity of blocking over the vectors representing the tuples.


\stitle{Contributions.} 
In this paper, we present \sys, a novel ER system powered by \DRs of tuples, which is accurate, efficient, and easy-to-use.
%

\stitle{1. [Distributed representations of tuples for ER.]} 
We present two methods for effectively computing \DRs of tuples by composing the \DRs  of 
all the tokens within all attribute values of a tuple.
The first method is a simple averaging of the tokens' \DRs while the second uses
uni- and bi-directional recurrent neural networks (RNNs) with long short term memory (LSTM) hidden units to convert each tuple  to a \DR (\ie a vector).
%

\stitle{2. [Learning/tuning distributed representations.]}
We introduce an end-to-end approach to tune the \DRs that is customized for a specific ER task which improves the performance of \sys (Section~\ref{sec:e2e}).

\stitle{3. [Blocking for distributed representations.]}
We propose two efficient and effective blocking algorithms based on the \DRs for tuples and locality sensitive hashing, 
which takes the semantic relatedness of all attributes into account (Section~\ref{sec:blocking}).
	
\stitle{4. [Experiments.]} 
We conducted extensive experiments (Section~\ref{sec:expt}). 
\sys shows superior performance compared to a state-of-the-art ER solution as well to published methods on several benchmark datasets from citations, products, and proteomics. 
Finally, the proposed blocking delivers outstanding results under different conditions.
\section{Distributed Representations of Tuples for Entity Resolution}
\label{sec:distrRepr}


\subsection{Entity Resolution}
\label{subsec:er}

Let $T$ be a set of entities with $n$ tuples and $m$ attributes $\{A_1, \ldots, A_m\}$.
Note that these entities can come from one table or multiple tables (with aligned attributes).
We denote by $t[A_i]$ the value of attribute $A_i$ on tuple $t$.
The problem of {\em entity resolution} (ER) is, given all distinct tuple pairs $(t, t')$ from $T$ where $t \neq t'$, to determine which pairs of tuples refer to the same real-world entities (\aka~a {\em match}).

\subsection{Distributed Representations of Words}
\label{subsec:weud}

\begin{figure}[t!]
	\vspace*{-2ex}
	\centering
	\includegraphics[width=0.8\columnwidth]{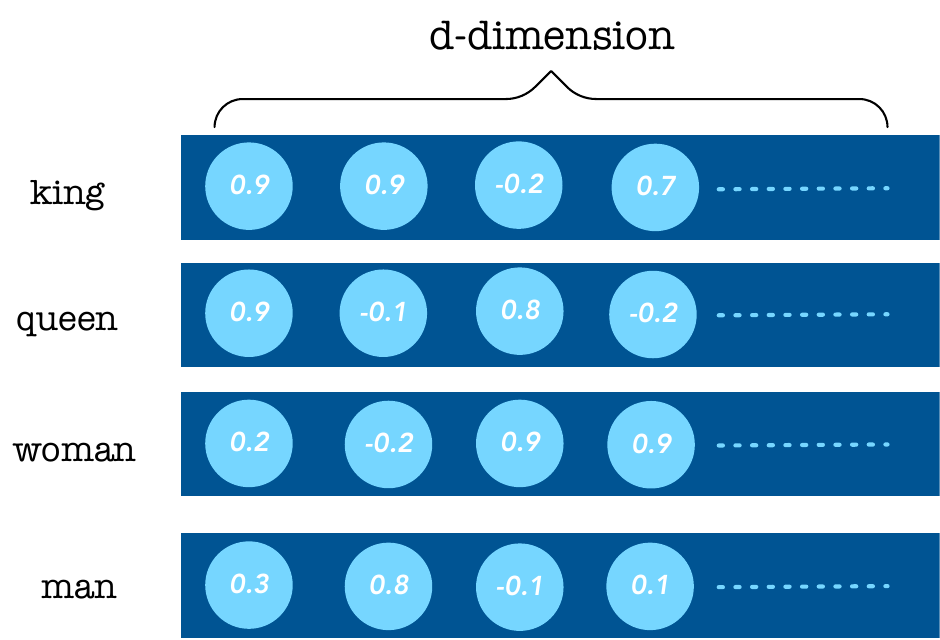}
	\vspace*{-2ex}
	\caption{Sample Word Embeddings}
	\label{fig:we}
	\vspace*{-3ex}	
\end{figure}

We briefly describe the concept of distributed representations (\DRs) of words (please refer to~\cite{Goodfellow-et-al-2016} for more details).
{\em \DRs of words} (\aka {\em word embeddings}) are learned from the data in such a way that semantically related words are often close to each other;
\ie the geometric relationship between words often also encode a semantic relationship between them.
This embedding method seeks to map each word in a given vocabulary into a high dimensional vector with a fixed dimension $d$ (\eg $d=300$ for GloVe).
In other words, each word is represented as a distribution of weights (positive or negative) across these dimensions. 
Figure~\ref{fig:we} shows some sample word embeddings.

Often, many of these dimensions can be independently varied.
The representation is  considered ``distributed'' since each word is represented by setting appropriate weights over multiple dimensions
while each dimension of the vector contributes to the representation of many words.
\DRs can express an exponential number of ``concepts''
due to the ability to compose the activation of many dimensions~\cite{Goodfellow-et-al-2016}.
In contrast, the symbolic (\aka discrete) representation often leads to data sparsity and requires substantially more data to train ML models successfully~\cite{Bengio03}.
Word embeddings have been successfully used to solve various tasks such as 
topic detection, document classification, and named entity recognition. 

A number of methods have been proposed to compute the \DRs of words including word2vec~\cite{mikolov2013distributed}, GloVe~\cite{pennington2014glove}, and fastText~\cite{bojanowski2016enriching}.
Generally speaking, these approaches attempt to capture the semantics of a word by considering its relations with neighboring words in its context. 
In this paper, we use GloVe, which is based on a key observation that the ratios of co-occurrence probabilities for a pair of words have some potential to encode a notion of its meaning. 
GloVe formalizes this observation as a log-bilinear model with a weighted least-squares objective function.
This objective function has a number of appealing properties such as 
the vector difference between the representations for (man, woman) and (king, queen) are roughly equal.


\subsection{Distributed Representations of Tuples} 
\label{subsec:wet}

Similar to word embeddings, given a tuple, we need to convert it to a vector representation, such that we can measure the similarity between two tuples by 
computing the distance between their corresponding vectors.
%

\begin{algorithm}[!t]
     \caption{A Simple Averaging Approach}
     \label{alg:tuple2VecSimple}
     \begin{algorithmic}[1]
        \STATE {\bf Input:} Tuple $t$, a pre-trained dictionary such as GloVe
        \STATE {\bf Output:} Distributed representation $\mathbf{v}(t)$ for $t$
        \FOR   {each attribute $A_k$ of $t$}
            \STATE Tokenize $t[A_k]$ into a set of words $W$
            \STATE Look up vectors for tokens $w_l \in W$ in GloVe
            \STATE $\mathbf{v}_k(t)$ := average of vectors of tokens in $t[A_k]$
        \ENDFOR
        \STATE ${\bf v}(t)$ := concatenation of ${\bf v}(t[A_k])$, for $k \in [1, m]$
     \end{algorithmic}
 \end{algorithm}

Consider a tuple $t$ with $m$ attributes $\{A_1, \ldots, A_m\}$. Let $\mathbf{v}(t[A_k])$ be the vector representation of value $t[A_k]$, and $\mathbf{v}(t)$ be the vector representation of tuple $t$. Also, we write $\mathbf{v}(x)$ the vector representation of a word $x$.
We also write $|{\bf v}|$ as the number of dimensions of vector ${\bf v}$.

\stitle{Running Example.}
We use the following example to illustrate our approaches. 
Table~\ref{tbl:runningEgTable} provides a toy relation with 2 tuples and Table~\ref{tbl:runningEgDR} provides word embeddings with $3$ dimensions.

\begin{table}[h!]
    \begin{center}
        \tabcaption{Toy Dataset for ER}
    \label{tbl:runningEgTable}
    \begin{tabular}{|c|c|c|}
        \hline
        {\bf Tuple ID} & {\bf $A_1$ : Name} & {\bf $A_2$ : City} \\ \hline
        $t_1$ & Bill Gates & Seattle \\ \hline
        $t_2$ & William Gates & Seattle \\ \hline
    \end{tabular}
    \end{center}
%
    \begin{center}
        \tabcaption{Sample Word Embeddings}
    \label{tbl:runningEgDR}
    \begin{tabular}{|c|c|}
        \hline
        {\bf Word} & {\bf Distributed Representation} \\ \hline
        Bill & $[0.4, 0.8, 0.9]$  \\ 
        William & $[0.3, 0.9, 0.7]$  \\
        Gates & $[0.5, 0.8, 0.8]$  \\
        Seattle & $[0.1, 0.1, 0.2]$  \\
        \hline
    \end{tabular}
    \end{center}
\end{table}

Below, we describe two approaches for computing $\mathbf{v}(t)$: a simple approach and a compositional approach. We then explain how we compute the similarity between  two vectors.

\subsubsection*{A Simple Approach -- Averaging}

For each attribute value $t[A_k]$, we first break it into individual words using a standard tokenizer.
For each token (word) $x$, we look up the GloVe pre-trained dictionary and retrieve the $d$-dimensional vector ${\bf v}(x)$.
If a word is not found in the GloVe dictionary (dubbed out-of-vocabulary scenario) 
or if the attribute has a NULL value, GloVe contains a special token {\sc{Unk}} to represent such out-of-vocabulary word (we postpone the discussion of a better handling of out-of-vocabulary values to Section~\ref{sec:e2e}).
Tokens such as IDs and some numeric values are often assigned {\sc {Unk}}.

In our initial approach, the vector representation for an attribute value ${\bf v}(t[A_k])$ is obtained by simply \emph{averaging} the vectors of its tokens $x$ in $t[A_k]$. 
The vector representation ${\bf v}(t)$ of tuple $t$ is the concatenation of all vectors ${\bf v}(t[A_k])$ ($k \in [1, m]$).
That is, if each attribute value corresponds to a $d$-dimensional vector, $|{\bf v}(t)| = d \times m$.
Algorithm \ref{alg:tuple2VecSimple} describes this process.

\begin{algorithm}[t]
     \caption{A Compositional Approach}
     \label{alg:tuple2VecCompositional}
     \begin{algorithmic}[1]
        \STATE {\bf Input:} Tuple $t$, a pre-trained dictionary such as GloVe
        \STATE {\bf Output:} Distributed representation $\mathbf{v}(t)$ for $t$
        \FOR   {each attribute $A_k$ ($k \in [1, m]$) of $t$}
            \STATE Tokenize $t[A_k]$ into a set of words $W$
            \STATE Look up vectors for tokens $w_l \in W$ in GloVe
 	        \STATE Pass the GloVe vectors for tokens through a LSTM-RNN composer to obtain $\mathbf{v}(t[A_k])$
        \ENDFOR
	\STATE ${\bf v}(t) := {\bf v}(t[A_m])$
    \end{algorithmic}
 \end{algorithm}

\begin{figure}[t!]
	\vspace*{-2ex}
	\centering
	\includegraphics[width=\columnwidth]{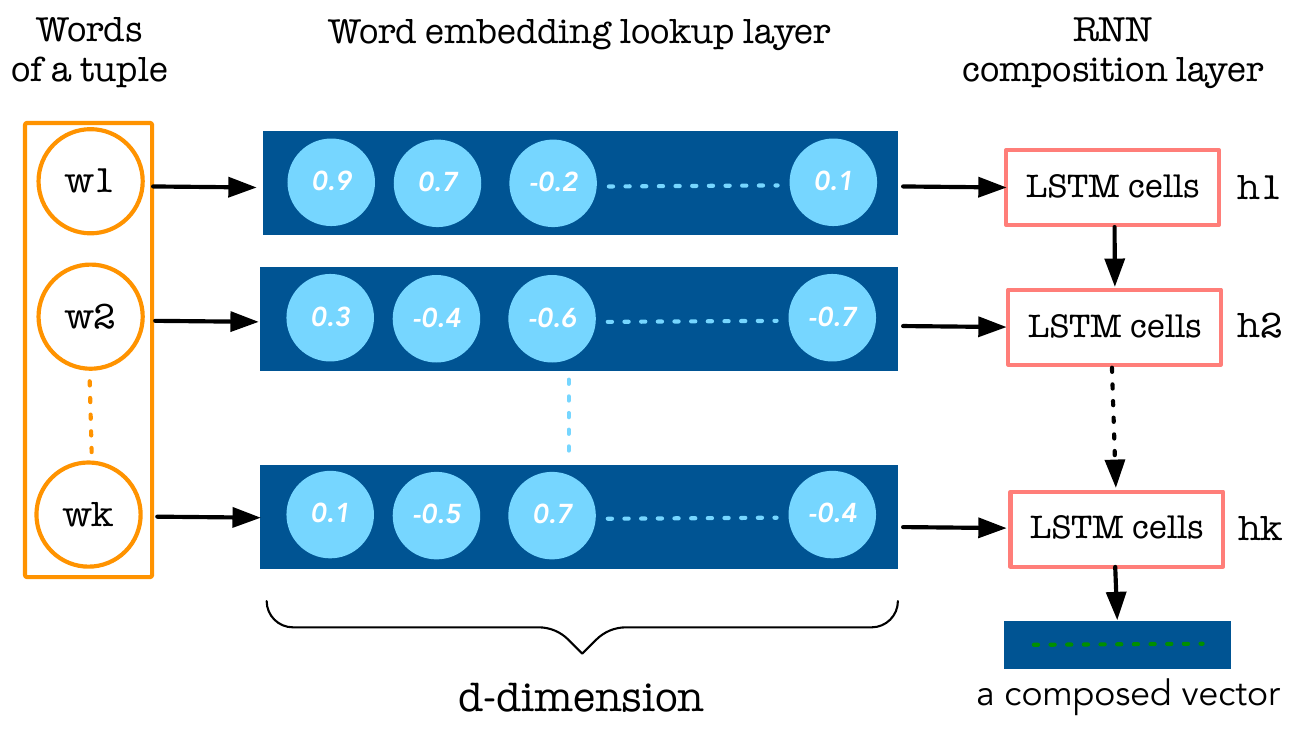}
	\vspace*{-4ex}
	\caption{RNN with LSTM in the Hidden Layer}
	\label{fig:lstmrnn}
	\vspace*{-2ex}	
\end{figure}

\begin{example}    
\label{exam:avg}
Using our running example, $v_1[t_1] = [0.45, 0.8,  0.85]$ and $v_1[t_2] = [0.4,  0.85, 0.75]$.
    $v_2[t_1] = v_2[t_2]$ = $[0.1, 0.1, 0.2]$.
    The \DR for $t_1$ and $t_2$ are obtained by concatenating the \DRs for $A_1$ and $A_2$.
\vspace{-1ex}
\end{example}


\subsubsection*{A Compositional Approach -- RNN with LSTM}\label{subsec:lstm}
We can see that the averaging based approach ignores the word order.
However, it has the appealing property of being simple and very efficient to train.
An alternative approach for computing $\mathbf{v}(t)$ is to use a compositional technique motivated by the linguistic structures (\eg $n$-grams, sequence, and tree) in Natural Language Processing (NLP). In this approach, instead of simple averaging, we use a neural network to semantically compose the word vectors (retrieved from GloVe) into an attribute-level vector. 
Considering the word order (and linguistic structure in general)  could be important as many attributes
contain multi-word content such as title in the citation dataset and description in the product dataset.
As we shall show in our experiments, there are certain challenging datasets where being cognizant of word order improves performance.
Furthermore, a tuple binds the attributes of a single entity, thus the attributes are related. 
Thus, an appropriate compositional approach should consider the relation between them rather than treating them separately.
Different neural network architectures have been proposed to consider different types of linguistic structures,
the most popular of which use a recurrent structure \cite{li-EtAl:2015:EMNLP5}. 

We use uni- and bi-directional recurrent neural networks (RNN) with long short term memory (LSTM) hidden units \cite{Hochreiter:1997}, \aka LSTM-RNNs. 
As shown in Figure~\ref{fig:lstmrnn}, RNNs encode a sequence of words for all attribute values (\ie words of a tuple) into a composed vector by processing its word vectors sequentially (\ie the word embedding lookup layer),
at each time step, combining the current input word vector with the previous hidden state (\ie the RNN composition layer). 
The outputted composed vector of ${\bf v}[t]$ has $x$ dimensions, where $x$ is determined by LSTM and may be different than $d$.
RNNs thus create internal states by remembering the output of the previous time step, 
which allows them to exhibit dynamic temporal behavior. 
We can interpret the hidden state $\mathbf{h}_i$ at time $i$ as an intermediate representation summarizing the past. The output of the last time step ${\mathbf{h}_k}$ thus represents the tuple. LSTM cells contain specifically designed gates to store, modify or erase information, which allow RNNs to learn long range sequential dependencies. The LSTM-RNN shown in Figure~\ref{fig:lstmrnn} is unidirectional in the sense that it encodes information from left to right. 

Bidirectional RNNs \cite{Schuster:1997} capture dependencies from both directions, thus provide two different views of the same sequence. For bidirectional RNNs, we use the concatenated vector $[\overrightarrow{\mathbf{h}_k},\overleftarrow{\mathbf{h}_k}]$ as the final representation of the attribute value, where $\overrightarrow{\mathbf{h}_k}$ and $\overleftarrow{\mathbf{h}_k}$ are the encoded vectors from left to right and right to left, respectively. 

Algorithm~\ref{alg:tuple2VecCompositional} gives the overall compositional process. For each word token in an attribute, we first look up its GloVe vector. 
Then we use a ``shared''\footnote{By the term `shared' we mean the parameters of the model are shared across the attributes. In other words, the LSTM-RNNs for different attributes in a table share the same parameters.}  LSTM-RNN to compose each attribute value in a tuple into a vector.
This results in a vector ${\bf v}(t)$ of $d$ dimensions.
It is important to note that the parameters of the LSTM-RNN model need to be learned on the ER task in a DL framework before it can be used to compose vectors for other off-the-shelf classifiers.  

The order of attributes might not have a big effect when the number of attributes is small (such as 4 for the citations benchmark datasets).
However, it might become significant when the number of attributes is large. 
In this case, a simple heuristic would be to ensure that semantically related attributes are close to each other. 
This is performed by profiling the data to find possible data dependencies. 
For example, if one identified that two attributes $A_i$ and $A_j$ are closely related 
(e.g. Country often determines Capital), then these can be ordered so that they are closer to each other.

\begin{example}
\label{exam:lstm}
Assume that we used a LSTM with output dimension of 2.
    In other words, it processes the entire tuple and produces a \DR of dimension $2$.
    Assume that $v(t_1) = [0.45, 0.23]$ and $v(t_2) = [0.42, 0.28]$. 
\vspace{-1.2ex}
\end{example}

\subsubsection*{Computing Distributional Similarity}

Given the \DR of a pair of tuples $t$ and $t'$, 
the next step is to compute the similarity between their \DRs ${\bf v}(t)$ and ${\bf v}(t')$.
Note that for the \DRs computed by averaging, each vector has $d\times m$ dimensions.  
We apply the cosine similarity on every $d$ dimensions (each $d$ dimension corresponds to one attribute), which results in a $m$-dimensional similarity vector. 
For the \DRs computed by LSTM, each vector has $x$ dimensions, we can use  methods including 
subtracting (vector difference) or multiplying (hadamard product) the corresponding vector entries, resulting in a $x$-dimensional similarity vector.

\begin{example}
\label{exam:ds}
    Continuing our running example, the similarity vector for tuples $t_1$ and $t_2$ for averaging is $[0.99, 1.0]$.
    The first component corresponds to cosine similarity of name and second the city.
    The distributional similarity for LSTM with vector differencing is $[0.03, -0.05]$.
\vspace{-1.2ex}
\end{example}

%


%
 
\begin{algorithm}[t!]
     \caption{ER -- Classifier}
     \label{alg:ERClassifier}
     \begin{algorithmic}[1]
        \STATE {\bf Input:} Table $T$, training set $S$ 
	   \STATE {\bf Output:} All matching tuple pairs in table $T$
	\STATE \add{/\!/ Training}
        \FOR   {each pair of tuples $(t, t')$ in $S$}
            \STATE Compute the distributed representation for $t$ and $t'$
            \STATE Compute their distributional similarity vector 
        \ENDFOR
        \STATE Train a classifier $\mathcal{C}$ using the similarity vectors for $S$ and true labels
        \STATE \add{/\!/ Testing}
        \FOR {each pair of tuples $(t, t')$ in $T$}
            \STATE Compute the distributed representation for $t$ and $t'$
            \STATE Compute their distributional similarity vector 
            \STATE Predict match/mismatch for $(t, t')$ using $\mathcal{C}$
        \ENDFOR
     \end{algorithmic}
 \end{algorithm}

\subsection{ER as a Classification Problem using Distributed Representations of Tuples}
\label{subsec:ercp}


ER is typically treated as a binary classification problem~\cite{theoryrecordlinkage,DBLP:journals/tkde/ElmagarmidIV07,naumann2010introduction,DBLP:conf/kdd/BilenkoM03}.
%
Given a pair of tuples $(t, t')$, the classifier outputs \True~(resp. \False) 
to indicate that $t$ and $t'$ match (resp. mismatch).
The Fellegi-Sunter model~\cite{theoryrecordlinkage} is a formal framework for probabilistic ER 
and most prior machine learning (ML) works are simple variants of this approach.

Intuitively, given two tuples $t$ and $t'$, we compute a set of similarity scores between aligned attributes based on predefined similarity functions. 
The vectors of known matching (resp. non-matching) tuple pairs -- that are also referred to as positive (resp. negative) examples -- 
are used to train a binary classifier (\eg SVMs,  decision trees, or random forests). 
The trained binary classifier can then be used to predict whether an arbitrary tuple pair is a match.

It is fairly straightforward to build a classifier for ER using the above steps.
For each  pair of tuples in the training dataset, we compute their \DRs 
through either Algorithm~\ref{alg:tuple2VecSimple} or Algorithm~\ref{alg:tuple2VecCompositional}.
We then compute the similarity between tuple pairs using different metrics. 
Given a set of positive and negative matching examples, we pass their similarity vectors to a classifier such as SVM along with their labels.  
Algorithm~\ref{alg:ERClassifier} provides the pseudocode.

\section{Learning and Tuning \\Distributed Representations}
\label{sec:e2e}

Composing \DRs of words to generate \DRs of tuples, as discussed in Section~\ref{sec:distrRepr}, works effectively for an ER task based on two assumptions:
(i) there exist pre-trained word embeddings for most (if not all) words in the dataset; and
(ii) the pre-trained word embeddings that were trained in a task agnostic manner are sufficient for the ER task.
However, the above two assumptions may not hold for many real-world scenarios.
%
The datasets that follow the above assumptions are considered as
{\em general data with full coverage} (Section~\ref{subsec:fullcover});
the datasets that are not well covered, are considered as 
{\em general data with partial coverage} (Section~\ref{subsec:substantialcover}); and
the datasets that are minimally covered, are considered as 
{\em specific data with minimal coverage} (Section~\ref{subsec:minimalcover}).
Finally, we discuss how to {\em tune word embeddings} for an ER task (Section~\ref{subsec:tunewe}).


\subsection{General Data with Full Coverage}
\label{subsec:fullcover}

Many of the benchmark datasets used in ER~\cite{magellandata} such as Citations, Products, Restaurants, and Movies,
are often generic and do not require substantial specialized knowledge.
While they may be noisy and incomplete, the content is often in English and use common words.
For such generic datasets, the approach that we have proposed so far -- convert pairs of tuples to similarity vectors using GloVe -- is often adequate.
As we shall show in the experiments, we obtain competitive results for all of them.

\subsection{General Data with Partial Coverage}
\label{subsec:substantialcover}

Another case is where a significant number of words that are relevant for an ER task in a dataset are not present in the word embedding dictionary.
It is well known that natural languages exhibit a Zipfian distribution with a heavy tail of infrequently used words.
For computational efficiency, these ``rare'' words are often pruned. 
For example, even though GloVe was trained on a very large document corpus with 840 billion tokens and a vocabulary size of 2.2 million,
it can miss many words such as from technical domains, names of people, or institutions, 
which would be useful for performing ER on a Citation dataset.
Our approach from Section~\ref{sec:distrRepr} replaces any word not present in the dictionary with a unique token UNK (Unknown).
However, it is possible that these words are especially relevant for identifying duplicate tuples.

\stitle{Vocabulary Expansion} is the process of expanding the embedding dictionary to words that were not observed during training.
The naive approach of adding new documents to the original corpus and re-run the whole algorithm is not always possible or feasible.
For example, the popular GloVe dictionary is trained on the Common Crawl corpus, which is almost 2 TB requiring exorbitant computing resources.
Given an unseen word, another simplistic approach is to take the top-$K$ words that co-occur the most with the unseen word in the ER dataset and simply average them.
Another popular approach is to use character level embeddings such as fastText instead of word level embeddings or use subword information~\cite{bojanowski2016enriching}.
These approaches can recognize that the rare word ``dendritic'' is similar to ``dendrites'' even if it is not explicitly present in the dictionary.

While these approaches are simple and often effective, in this paper, we advocate an alternative approach inspired from~\cite{faruqui2014retrofitting}, as described below.

\stitle{Vocabulary Retrofitting.}
Intuitively, this approach seeks to adapt word embeddings such as GloVe by using auxiliary semantic resources such as WordNet.
If there are two words that are related in WordNet, ~\cite{faruqui2014retrofitting} seeks to refine their word embeddings to be similar.
This approach is especially relevant to our scenario where our input is a tuple with explicit attribute structure and has relational interpretations. 

\begin{figure}[t!]
	\centering
	\includegraphics[width=0.6\columnwidth]{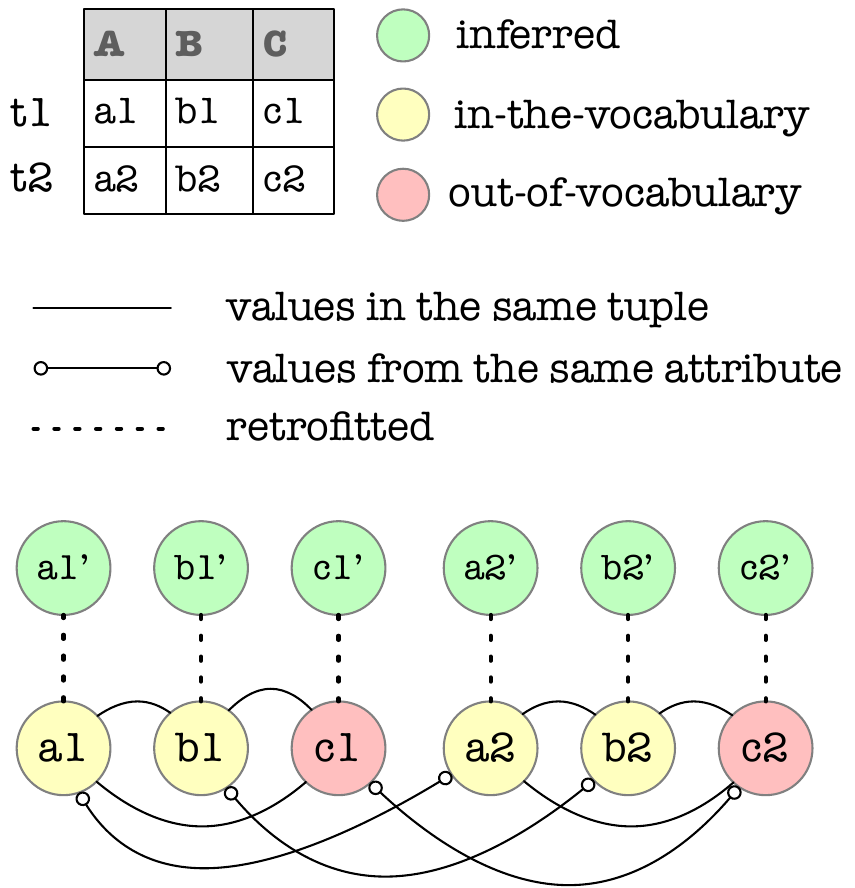}
	\vspace*{-2ex}	
	\caption{A Sample for Vocabulary Retrofitting}
	\label{fig:vr}
	\vspace*{-2ex}	
\end{figure}

Let $W = \{w_1, w_2, \ldots\}$ be the set of words from the ER dataset. 
Let $U \subseteq W$ be the set of words with no word embeddings.
We begin by creating an undirected graph with one vertex for each word in $W$.
Two vertices $(v_i, v_j)$ are connected if they co-occur in some tuple.
One can also define other additional edge semantics such as an edge connecting two vertices if they occur in the same attribute and so on.
For vertex $v \in W \setminus U$, we assign its word embedding from the dictionary.
For vertex $v \in U$, we assign its initial value as the average of $K$ of its most frequent co-occurring words.
We then create a set of new vertices $W^\prime$ for each word $w \in W$ and connect it to its corresponding vertex from $W$.
These vertices correspond to the retrofitted word embeddings that will be learned through probabilistic inference.
We set the objective function such that we learn the word embeddings for each vertex $w \in W^\prime$ 
such that they are both closer to its counterpart in $W$ and also closer to its other neighboring vertices.
Figure~\ref{fig:vr} shows an example.

This approach has a number of appealing properties.
First, this is an efficient mechanism to learn the word embeddings of unknown words such as IDs.
Second, it provides an elegant mechanism to ``tune'' the word embeddings that is in sync with the ER dataset.
For example, if two words do not co-occur a lot in Common Crawl (such as SIGMOD and Stonebraker) but does in our dataset, this approach tunes the embeddings appropriately.
Finally, it allows one to encode a diverse array of options to define relatedness (such as present in the same attribute or tuple) 
that is more generic than the simple co-occurrence idea used by GloVe.

\subsection{Specific Data with Minimal Coverage}
\label{subsec:minimalcover}

Another common scenario occurs when one performs ER on datasets with specialized information.
Examples include performing ER on scientific articles for specialized fields or in data that is specific to an organization.
In this case, the attributes often contain esoteric words that are not present in GloVe dictionary.
What is worse, it might not know that two terms such as p53 and cancer are related.
If most of the words are not present in the dictionary, then approaches such as retrofitting might be applicable.
This problem could be exacerbated if the ER dataset belongs to complex domains such as Genomics.
Finding if two tuples describe the same protein might not be possible with GloVe or word2vec.

We use one of the following approaches to handle such scenarios:

\etitle{(1) Unsupervised Representation from Datasets.}
        If the two datasets that are to be merged are large enough (typically in the order of millions of tuples), 
        they might contain enough patterns to automatically learn word embeddings from them.
        One could pool all the tuples from both datasets and train GloVe/word2vec on them by treating each tuple as a document.
        
\etitle{(2) Unsupervised Learning from related Corpus.}
        If the datasets are not large enough, then one can find another surrogate resource to learn word embeddings.
        GloVe and word2vec learned the word embeddings by training on a large corpus such as Common Crawl and Google News respectively.
        If one can find an analogous vast corpus of domain information in the form of unstructured data such as documents,
        it could be used to learn the word embeddings for this specialized dataset.
        For example, while word embeddings from GloVe might not know that p53 and cancer are related,
        the word embeddings trained from PubMed articles would be able to.
        Similarly, one could learn word embeddings from the enterprise's document repository for ER on data in the same organization.

\etitle{(3) Customized Word Embeddings.}
        In some cases, it is possible that direct application of GloVe or word2vec does not solve the problem.
        For example, when given a huge amount of training they might not find if, for example, two strings encode the same protein.
        In such cases, one has to check if there exist prior methods for learning word embeddings for the task of interest. 
        For example, there exist prior work on word embeddings for proteins and genes from sequences of amino acids and DNA respectively~\cite{asgari2015continuous}.

Of course, the worst case scenario is a specialized database where no 
auxiliary resources are available to automatically learn the representation for key concepts.
In this scenario, any machine learning approach is doomed to fail unless one provides 
hand crafted features or a substantially large number of training examples that are sufficient for learning representations using deep learning.

\begin{figure}[t!]
	\centering
	\includegraphics[width=0.9\columnwidth]{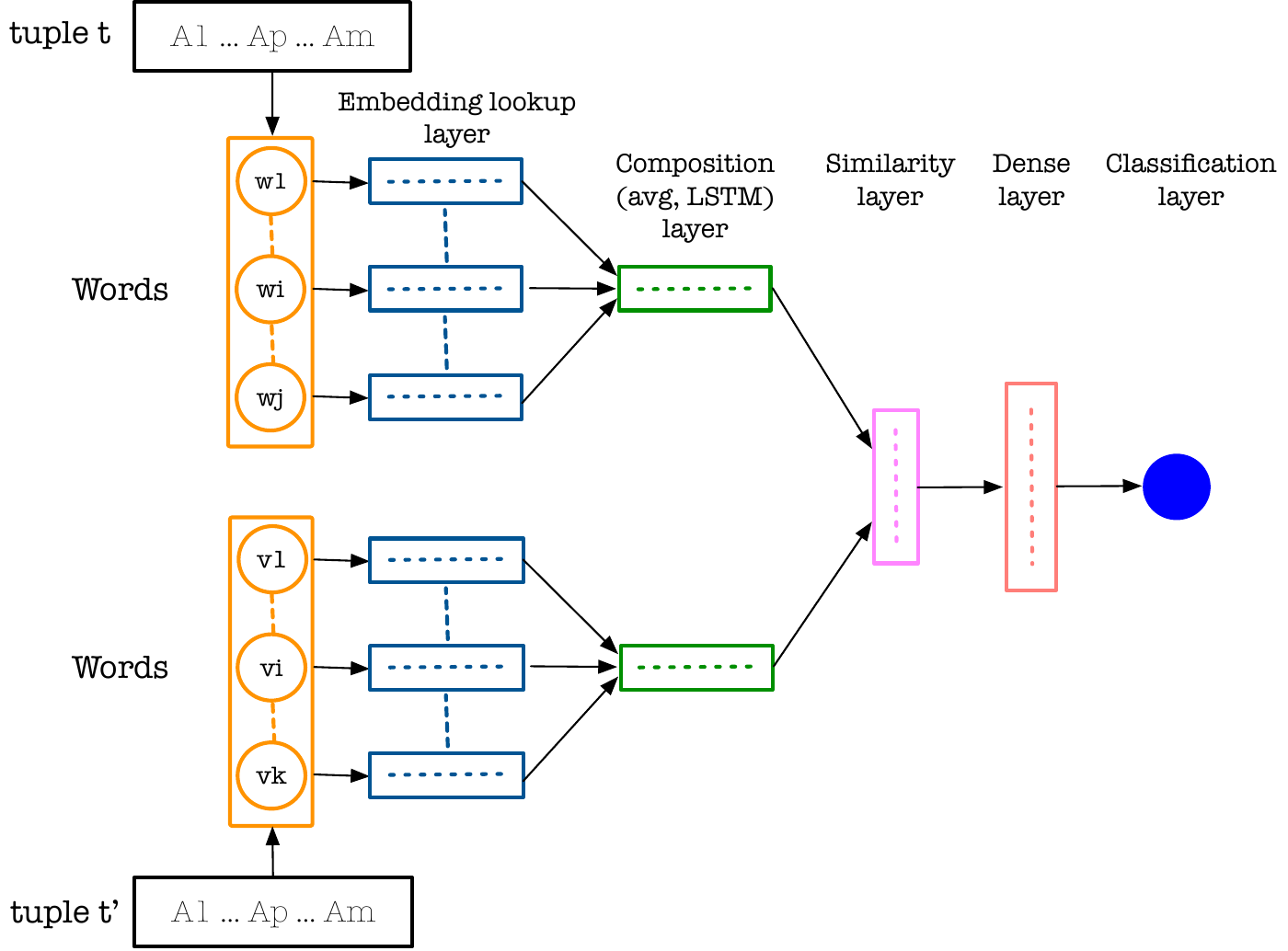}
	\vspace*{-2ex}		
	\caption{Deep Entity Resolution Framework}
	\label{fig:ERmodel}
	\vspace*{-2ex}	
\end{figure}

\subsection{Tuning Word Embeddings for an ER Task}
\label{subsec:tunewe}

Recall that word embeddings such as GloVe/word2vec are learned in an unsupervised task-agnostic manner so that they can be used for arbitrary tasks.
If the corpus used to train them is large and representative enough, the learned word embeddings can be used in a turn-key manner for ER tasks. 
While unsupervised pre-training on a large corpus does give the DL model better generalization,
in many cases the learned representations often lack task-specific knowledge.
One can achieve the best of both worlds by fine-tuning the pre-trained word representations to achieve better accuracy. 
In fact, this paradigm of unsupervised pre-training followed by supervised fine-tuning often beats methods that are based on only supervision~\cite{collobert2011natural}.

Our proposed approach can be easily extended for this purpose. 
Let us now consider our deep neural network in Figure~\ref{fig:ERmodel}. 
We train this network using Stochastic Gradient Descent (SGD) based learning algorithms, where gradients (errors) are obtained via backpropagation. 
In other words, errors in the output layer (\ie the classification layer) are backpropagated through the hidden layers using the chain rule of derivatives. 
The parameters of the hidden layers are slightly altered such that when the model accuracy improves.
For learning or fine-tuning the embeddings, we allow these errors to be backpropagated all the way till the word embedding layer. 
In contrast, our approach from Section~\ref{sec:distrRepr} can only tune parameters up to the composition layer.
Allowing error to be back propagated to the embedding layer allows one additional level of freedom to tinker model parameters.
Instead of limiting ourselves to how the attributes are composed or how similarity is computed, 
we can also modify the word embeddings themselves (if necessary).

One common issue with backpropagation through a deep neural network (\ie neural networks with many hidden layers such as  RNNs) is that as the errors get propagated, they may soon become very small (\aka gradient vanishing problem)  or very large (\aka gradient exploding problem) that can lead to undesired values in weight matrices, causing the training to fail \cite{Bengio:1994}. We did not observe such problems in our end-to-end training  with simple averaging compositional method, and the gates in LSTM cells automatically tackle these issues to some extent \cite{Hochreiter:1997}.

\section{Blocking for Distributed Representations}
\label{sec:blocking}

Efficient ER systems avoid comparing all possible pairs of tuples 
(${n \choose 2}$  for one table of $n \times m$ for two tables) through the use of blocking~\cite{blocking,christen2012survey}. 
Blocking identifies groups of tuples (called {\em blocks}) such that the search for duplicates need to be done only within these blocks, thus greatly reducing the search space.
While blocking often substantially reduces the number of comparisons, it may also miss some duplicates that fall in two different blocks.

\subsection{New Opportunities for Blocking}
We observe that blocking is very related to the classical problem of approximate nearest neighbor (ANN) search in a similarity space, 
which has been extensively studied (see \cite{wang2014hashing}).
Locality sensitive hashing (LSH)~\cite{DBLP:conf/stoc/IndykM98} is a popular probabilistic technique for finding ANNs in a high dimensional space.
In the blocking context, the more similar input vectors are, 
the higher the probability that they both will be put in the same block.
While we are not the first to propose LSH for blocking or automated tuning for blocking (see Section~\ref{sec:related}),
we are the first to propose a series of truly turn-key algorithms for blocking.

\subsubsection*{Challenges in Traditional Blocking Approaches}

\sstab
(i) Identifying good blocking rules often requires the assistance of domain experts.
   
\sstab
(ii) Blocking rules often consider few (\eg 2-3) attributes which could result in comparing tuples that agree on those attributes but have very different values on other attributes.

\sstab
(iii) Prior blocking methods often do not take semantic similarity between tuples into consideration.

\sstab
(iv) It is usually hard to tune the blocking strategy to control the recall and/or the size of the blocks.

We can readily see that LSH for blocking over \DRs of tuples obviates many of these issues.
First, we free the domain experts from providing a blocking function. 
Instead the combination of LSH and \DRs transforms the problem of blocking into finding tuples in a high dimensional similarity space.
Note that a \DR encodes semantic similarity into the mix and that LSH considers the entire tuple for computing similarity.
The extensive amount of theoretical work on LSH (see Section~\ref{subsec:LSHTuning}) 
can be used to both tune and provide rigorous theoretical guarantees on the performance.

\subsection{LSH Primer}
\label{fig:lshprimer}
\begin{definition}
    \label{def:lshFamily}
    ({\em Locality Sensitive Hashing}~\cite{gionis1999similarity,wang2014hashing}): 
    A family $\mathcal{H}$ of hash functions is called $(R, cR, P_1, P_2)$-sensitive if for any two items $p$ and $q$,
    \begin{itemize}
        \item if $dist(p,q) \leq R$, then $Prob[h(p) = h(q)] \geq P_1$, and
        \item if $dist(p,q) \geq cR$, then $Prob[h(p) = h(q)] \leq P_2$,
    \end{itemize}
    where $c > 1$, $P_1 > P_2$, $h \in \mathcal{H}$. 
\end{definition}

The smaller the value of $\rho$ ($\rho = \frac{\log(1/P_1)}{\log(1/P_2)}$), the better the search performance. 
For many popular distance measures such as cosine, Euclidean, and Jaccard,
there exists an algorithm for the $(R,c)$-nearest neighbor problem 
that requires $O(dn + n^{1 + \rho})$ space (where $d$ is the dimensionality of $p, q$), 
$O(n^\rho)$ query time,  and 
$O(n^\rho \log_{1/P_2} n)$ invocations of hash functions.
In practice, LSH requires linear space and time \cite{gionis1999similarity,wang2014hashing}.

\stitle{Implementing LSH.} 
Given a table $T$, LSH seeks to index all the tuples in a hash table
that is composed of multiple buckets each of which is identified by a unique hash code.
Given a tuple $t$, the bucket in which it is placed by a (single) hash function $h$ is denoted as $h(t)$ - which is often a binary value.
If two tuples $t$ and $t^\prime$ are very similar, then $h(t) = h(t^\prime)$ with high probability.
Typically, one uses $K$ hash functions $h_1(t), h_2(t), \ldots, h_K(t)$, $h_i \in \mathcal{H}$, 
to encode a single tuple $t$.
We represent $t$ as a $K$ dimensional binary vector 
which in turn is represented by its hash code $g(t) = (h_1(t), h_2(t), \ldots, h_K(t))$.
Since the usage of $K$ hash functions reduces the likelihood that similar items will obtain the same ($K$ dimensional) hash code,
we repeat the above process $L$ times - $g_1(t), g_2(t), \ldots, g_L(t)$.
Intuitively, we build $L$ hash tables where each bucket in a hash table is represented by a hash code of size $K$.
Each tuple is then hashed into $L$ different hash tables where its hash codes are $g_1(t), \ldots, g_L(t)$.
For example, if $K=10$ and $L=2$, every tuple is represented as a 10-dimensional binary vector that is stored in 2 different hash tables.

\stitle{Hash Families for Cosine Distance.}
Cosine similarity provides an effective method for measuring semantic similarity between two \DRs~\cite{pennington2014glove}.
Since the \DRs can have both  positive and negative real numbers,
the cosine similarity varies between $-1$ and $+1$.
The family of hash functions for cosine is obtained using the {\em random hyperplane} method.
Intuitively, we choose a random hyperplane through the origin that is defined by a normal unit vector $v$.
This defines a hash function with two buckets where $h(t) = +1$ if $v \cdot t \geq 0$ 
and $h(t) = -1$ if $v \cdot t < 0$ where ``$\cdot$'' denotes the dot product between vectors.
Since we require $K$ hash functions $h_1, \ldots, h_K$, we randomly pick $K$ hyperplanes 
and each tuple is hashed with them to obtain a $K$ dimensional hash code.
This process is then repeated for all $L$ hash tables.

\subsection{LSH-based Blocking}

We begin by generating hash codes $h_1, \ldots, h_K$ for each of the $L$ hash tables using the random hyperplane method.
The set of hash functions $h_1, \ldots, h_K$ is analogous to a single blocking rule.
The $K$ dimensional binary hash code is equivalent to an identifier to a distinct block where $t$ falls into.
Each hash table performs ``blocking'' using a different blocking rule.

We index the \DR of every tuple $t$ in each of the $L$ hash tables.
LSH guarantees that similar tuples get the same hash code (and hence fall into the same block) with high probability.
Then, we consider each of the blocks for every hash table and invoke the classifier over the distinct pairs of tuples found in them.
Algorithm~\ref{alg:blockingLSH} provides the corresponding pseudocode.

\begin{algorithm}[t]
     \caption{ER Classifier with LSH based Blocking}
     \label{alg:blockingLSH}
     \begin{algorithmic}[1]
        \STATE {\bf Input:} Table $T$, training set $S$, $L$ 
	   \STATE {\bf Output:} All matching tuple pairs in Table $T$
          \STATE Generate  hash functions for $g_1, \ldots, g_L$ using the random hyperplane method
            \FOR {each tuple $t$}
            \STATE Index the DR of $t$ into $L$ hash tables using $g_1, \ldots, g_L$
            \ENDFOR
            \FOR {each hash table $g$ in $[g_1, \ldots, g_L]$}
                \FOR {each non-empty bucket $H$ in $g$}
                    \FOR {each pair of tuples $(t, t')$ in $H$}
                        \STATE Apply classifier on $(t, t')$  
                    \ENDFOR
                \ENDFOR
            \ENDFOR
     \end{algorithmic}
 \end{algorithm}

\begin{example}
\label{exam:hash}
    For simplicity, let us only hash the \DR for attribute $A_1$ of tuples $t_1$ and $t_2$.
    Let $K=4$ and $L=1$.
    Let the hash functions be $h_1=[-1, 1, 1]$, $h_2=[1, 1, 1]$, $h_3=[-1, -1, 1]$ and $h_4=[-1, 1, -1]$.
    Recall that $v_1[t_1] = [0.45, 0.8,  0.85]$ and $v_1[t_2] = [0.4,  0.85, 0.75]$.
    If you do a dot product of $v_1[t_1]$ with each of the $h_i$'s, you get $[1.2, 2.1, -0.4, -0.5]$.
    The corresponding output for $v_1[t_2]$ is $[1.2, 2.0, -0.5, -0.3]$.
    Note that the LSH hash code is obtained by thresholding the values such that positive values get $+1$ and negative values $-1$.
    So the hash code of both these tuples is $[1, 1, -1, -1]$.
\vspace{-1.2ex}
\end{example}

Algorithm~\ref{alg:blockingLSH} is a fairly straightforward adaptation of LSH to ER.
As we shall see in the experiments, it works well empirically.
However, the number of times a classifier would be invoked 
can be as much as $O(L \times b^2_{max} \times B_{max})$
where $L$ is the number of hash tables, $b_{max}$ is the size of the largest block in any hash table and 
$B_{max}$ is the maximum number of non-empty blocks in any hash table.
While the traditional LSH based approach is often efficient and effective,
one can achieve improved performance with some additional domain knowledge.
We next describe a sophisticated approach to reduce the impact of $L$ and $b_{max}$.

\subsection{Multi-Probe LSH for Blocking}
\label{fig:multiprobelsh}
Recall that by increasing $K$, we ensure that the probability of dissimilar tuples falling into the same block is reduced.
By increasing $L$, we ensure that similar tuples fall into the same block in at least one of the $L$ hash tables.
Hence while increasing $L$ ensures that we will not miss a true duplicate pair, 
it is achieved at the additional cost of making extraneous comparisons between non-duplicate tuples.
We wish to come up with a LSH based approach that achieves two objectives:
(a)~reduce the number of unnecessary comparisons and
(b)~reduce the number of hash tables $L$ without seriously affecting recall.

\stitle{Reducing Unnecessary Comparisons.}
Intuitively, we expect duplicate tuples to have a high similarity with each other and thereby more likely to be ``near'' each other. 
Hence, even if a block has a large number of tuples, it is not necessary to compare all pairs of tuples.
Instead, given a tuple $t$, we retrieve the top-$N$ nearest neighbors of $t$ and invoke the classifier between $t$ and these $N$ nearest neighbors. 
This is achieved by collating all the tuples that fall into the same block as $t$ in each of the $L$ hash tables.
We then compute the similarity between $t$ and each of the candidates and return the top-$N$ tuples.
If the block is large with $b$ tuples, then we only require $\Theta(b \times N)$ classifier invocations instead of $\Theta(b^2)$.
We can see that by choosing $N < b$, we can achieve considerable reduction in classifier invocations.

\stitle{Reducing $L$.}
Naively decreasing the number of hash tables $L$ can decrease the recall as a pair of duplicate tuples might fall into different blocks.
The key idea is to augment a traditional LSH scheme with a carefully designed probing sequence that looks for 
multiple buckets (of the same hash table) that could contain similar tuples with high probability.
This approach is called  multi-probe LSH~\cite{lv2007multi}.
Consider a tuple $t$ and another very similar tuple $t^\prime$. 
It is possible that $t$ and $t^\prime$ do not fall into the same bucket (especially if $K$ is large).
However, due to the design of LSH, we would expect $t^\prime$ to fall into a ``close by'' bucket 
whose hash code is very similar to the bucket in which $t$ fell. 
Multi-probe leverages this observation by perturbing $t$ in a systematic manner and looking at all buckets in which the perturbed $t$ fell into.
By carefully designing the perturbation process one can consider the buckets that have the highest probability of containing similar tuples.
For example, a multi-probe of size 1 will consider all buckets whose hash codes have a hamming distance of 1 and so on.
It has been shown that this approach often requires substantially less number of hash tables (as much as 20x)
than a traditional approach \cite{lv2007multi}.
Algorithm~\ref{alg:lshANNSearch} provides the pseudocode of this approach.

\begin{algorithm}[t]
     \caption{Approximate Nearest Neighbor Blocking}
     \label{alg:lshANNSearch}
     \begin{algorithmic}[1]
            \STATE Index all tuples using LSH 
            \FOR {each tuple $t$}
                \STATE Get candidate tuples using Multiprobe-LSH
                \STATE Sort tuples in candidates based on similarity with $t$
                \STATE Invoke classifier on $t$ and each of top-$N$ neighbors of $t$
            \ENDFOR
     \end{algorithmic}
 \end{algorithm}

\subsection{Tuning LSH Parameters for Blocking}
\label{subsec:LSHTuning}

In contrast to traditional blocking rules that are often heuristics,
the hash functions in LSH allow us to provide rigorous theoretical guarantees.
While the list of LSH guarantees is beyond the scope of this paper (see \cite{steorts2014comparison} for details),
we highlight two major ones.

\stitle{Parameter Tuning for Recall.}
We can control the false positive and negative values (and thereby recall) by varying the values of $c$ and $R$,
such as by setting the values that get the best results for the tuples in the training dataset.
We can obtain a fixed approximation ratio of $c = 1 + \epsilon$ by setting~\cite{wang2014hashing},
\begin{equation} \label{eq:lshK} K = \frac{\log n}{\log 1/P_2} \quad
L = n^{\rho} \mbox{ where } \rho = \frac{\log(1/P_1)}{\log(1/P_2)} \end{equation}

\stitle{Parameter Tuning for Occupancy.}
LSH also allows us to control the occupancy - the expected number of tuples in any given block.
This can be achieved by varying the size $K$ of the number of hash functions in every hash table.
Informally, if one uses multiple hash functions, we would expect {\em very} similar items to 
be stored in the same blocks but at the expense of low occupancy and a large number of blocks.
On the other hand, a smaller number of hash functions results in less similar tuples being put in the same block.
Intuitively, if we use only one hash function, this results in 2 buckets - one for $+1$ and $-1$.
Since the hyperplane for the hash function is chosen randomly, 
we would expect each bucket to have an occupancy around 50\% for all but most of the skewed data distributions.
One can reduce the occupancy rate by increasing the number of hash functions.
Alternatively, one can also use sophisticated methods such as \cite{covell2009lsh} to achieve guaranteed limits.


\begin{table*}
    \begin{minipage}[t]{0.45\textwidth}
    \vspace{-1ex}
       {\small
    \begin{center}
        \caption{Data Statistics: $^*$(Easy),  $^\ddagger$(Hard). 
    Walmart-Amazon (Prod-WA),
    Amazon-Google (Prod-AG),
    DBLP-ACM (Pub-DA),
    DBLP-Scholar  (Pub-DS),
    DBLP-Citeseer (Pub-DC),
    Fodors-Zagat  (Rest-FZ)
    }
    \label{tbl:datasets}

        \begin{tabular}{|c|c|c|c|}
            \hline
            {\bf Dataset} & {\bf \#Tuples} & \hspace*{-1ex}{\bf \#Matches}\hspace*{-1ex} & \hspace*{-1ex}{\bf \#Attr}\hspace*{-1ex}\\ \hline
               \hspace*{-2ex} \shortstack{Walmart-Amazon$^\ddagger$\cite{magellandata}} \hspace*{-2ex} & 2,554 - 22,074 & 1,154 & 17 \\ \hline
                \shortstack{Amazon-Google$^\ddagger$~\cite{erhardws}} & 1,363 - 3,226 & 1,300 & 5 \\ \hline
                \shortstack{DBLP-ACM$^*$~\cite{erhardws}}  & 2,616 - 2,294 & 2,224 & 4 \\ \hline
                \shortstack{DBLP-Scholar$^*$~\cite{erhardws}} & 2,616 - 64,263 & 5,347 & 4 \\ \hline
                \shortstack{DBLP-Citeseer$^*$~\cite{magellandata}} & \hspace*{-1ex}1,823,978-2,512,927\hspace*{-1ex} & 558,787 & 4 \\ \hline
                \shortstack{Fodors-Zagat$^*$~\cite{uci}} & 533 - 331 & 112 & 7 \\ \hline
        \end{tabular}
    \end{center}
    } 
    \vspace{-3ex}
    \end{minipage}
    \hspace*{7ex}
    \begin{minipage}[t]{0.46\textwidth}
        {\small
    \begin{center}
    \caption{Comparing DeepER with state-of-the-art published results from existing rule-, ML- and crowd-based approaches. We also compared against Magellan, another end-to-end EM system.}
    \label{tbl:deepERBestPerf}
    
        \begin{tabular}{|c|c|c|c|}
            \hline
            {\bf Dataset} & {\bf Magellan} & {\bf DeepER} & {\bf Published}\\ \hline
            Prod-WA & 82.99& 88.06& 89.3~\cite{corleone}  (Crowd) \\ \hline
            Prod-AG & 87.68& 96.029& 62.2~\cite{erhard2010}  (ML) \\ \hline            
            Pub-DA & 97.6  & 98.6 & N/A \\ \hline
            Pub-DS & 98.84& 97.67& 92.1~\cite{corleone}  (Crowd) \\ \hline
            Pub-DC & 96.4 & 99.1 & 95.2~\cite{falcon} (Crowd) \\ \hline
            Rest-FZ & 100 & 100 & 96.5~\cite{corleone}  (Crowd) \\ \hline
        \end{tabular}
    \end{center}
    }
    \end{minipage}
       \vspace{-4ex} 
\end{table*}


\section{Experimental Results}
\label{sec:expt}

\subsection{Experimental Setup}
\label{sec:setting}

\stitle{Hardware and Platform.}
All our experiments were performed on a Core i7 6700HQ Skylake chip, with four cores running eight threads at speeds of up to 3.5GHz, along with 16GB of DDR4 RAM and the GTX 980M, complete with 8GB of DDR5 RAM. We used Torch~\cite{torch} and Keras~\cite{chollet2018deep}, a deep learning framework, to train and test our models. 

\stitle{Datasets.}
We conducted extensive experiments over $7$ different datasets covering diverse domains such as citations, e-commerce, and proteomics.
Table~\ref{tbl:datasets} provides some statistics of these datasets.
All are popular benchmark datasets and have been extensively evaluated by prior ER work using both ML and non-ML based approaches.
We partition our datasets into two categories: ``easy'' and ``challenging''.
The former consists of datasets that are mostly structured and often have less noise in terms of typos and missing information.
On the ``easy'' datasets most of the best existing ER approaches routinely exceed an F-score of $0.9$.
The challenging datasets often have unstructured attributes (such as product description) which are also noisy.
On the ``challenging'' datasets that we study, both ML and rule based methods have struggled to achieve high F measures, with values between $0.6$ and $0.7$ being the norm.
What these two categories have in common is that they require extensive effort from domain experts for cleaning, feature engineering and blocking to achieve good results.
As we shall show later, our approach exceeds best existing results on all the datasets with minimal expert effort.


\stitle{{\sys Setup.}}
Our experimental setup was an adaptation of prior ER evaluations methods \cite{kopcke2008training, erhard2010, synthesizer}
to handle \DRs.
For example, \cite{kopcke2008training} used an arbitrary threshold (such as 0.1) on Jaccard similarity of trigram to eliminate tuple pairs that are clearly non-matches.
We make two changes to this procedure.
First, we use Cosine similarity to compute similarity between tuple pairs as it is more appropriate for \DRs~\cite{mikolov2013distributed, pennington2014glove}.
Second, instead of picking an arbitrary threshold, we set it to the minimum similarity of matched tuple pairs in the {\em training} dataset.
We obtain the negative examples (non duplicates) by picking one tuple from the positive example and randomly picking another tuple  from the relation that is not its match.
For example, if $(t_i, t_j)$ is a duplicate, we pick a pair $(t_k, t_l)$ as a negative example 
such that $(t_k, t_l)$ is not a duplicate already given in the training data and has cosine similarity with $(t_i, t_j)$ below the above computed threshold.
This approach is chosen to verify the robustness of our models against near matches.
For each of the datasets, we performed $K$-fold cross validation with K=5.
We report the average of the F-measure values obtained across all the folds. 
{We observed that in all cases, the standard deviation of the F-measure values was below 1\%.}

\stitle{{\sys Architecture.}}
Since our objective is to highlight the turn-key aspect of \sys, we choose the simplest possible architecture.
We use GloVe\cite{pennington2014glove} as our \DR.
Figure~\ref{fig:ERmodel} shows the architecture of \sys.
We used Adam for optimizing the DL model that were trained for 20 epochs with a batch size of 16.
The learning rate was set to 0.01 and a regularization of 1e-3.
For RNN-LSTM composition, we used a single RNN layer where the memory dimension for LSTM is 150.
The dimension of the similarity layer is 50.
When end-to-end learning is enabled, the embeddings update rate was set to 0.01.

Each tuple is represented as a $m \times d$ dimensional vector where $m$ is the number of attributes with $d$ being the dimension of \DRs.
For each attribute, we apply a standard tokenizer and average the \DR obtained from GloVe (as against more sophisticated approaches such as Bi-LSTM).
Given a pair of tuples, the compositional similarity is computed as the Cosine similarity of the corresponding attributes resulting in a $m$ dimensional similarity vector.
As mentioned above, we used $K$-fold validation with a duplicate to non-duplicate ratio  of 1:100 that is comparable to the ratio used by the competing approaches. This is to ensure fair comparison.
Note that the non-duplicates are sampled automatically from the positive examples.
This is achieved by selecting other tuple pairs that have a low cosine similarity with the duplicate being perturbed.
This approach was inspired from~\cite{lazaridou2015hubness} where they demonstrate that an informative negative example is one
that is far from the positive example. 
Quoting from their example, ``truck'' is selected over ``dog'' as the negative example for ``cat''. 
We also do not tune the \DR for the ER task.
Even with this restricted setup, \sys is competitive with existing approaches.

\subsection{Comparison with Existing Methods}
\label{subsec:deepERVsPriorWork}

We first compare the performance of \sys with the best reported results from 
non-learning, learning and crowd based approaches in \cite{erhard2010, corleone,falcon}.
Table~\ref{tbl:deepERBestPerf} shows that
the simple \sys architecture has better performance, except for one dataset, namely Prod-WA, where a crowd-based approach does slightly better.



 \begin{figure*}[!t]
    \begin{minipage}[t]{0.33\linewidth}
      \centering
      \includegraphics[scale=0.45]{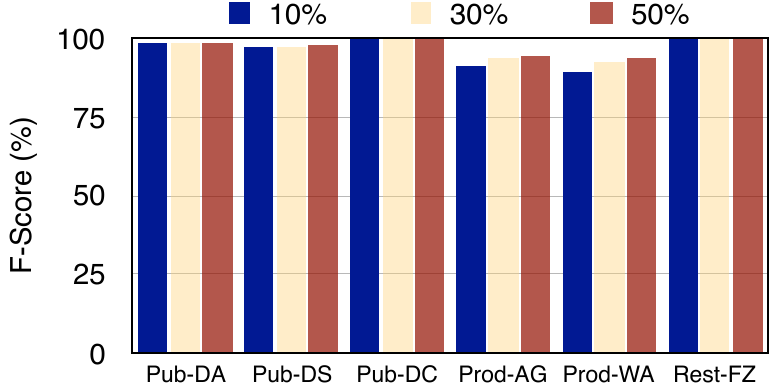}
      \vspace*{-4ex}
      \caption{Varying Training Data}
      \label{fig:varyingTrainingData}
    \end{minipage}
    \begin{minipage}[t]{0.33\linewidth}
      \centering
      \includegraphics[scale=0.45]{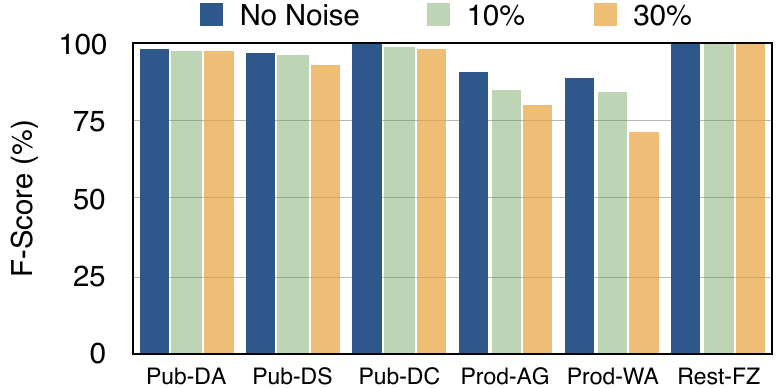}
      \vspace*{-4ex}
      \caption{Varying Noise}
      \label{fig:varyingNoise}
    \end{minipage}
    \begin{minipage}[t]{0.33\linewidth}
      \centering
      \includegraphics[scale=0.45]{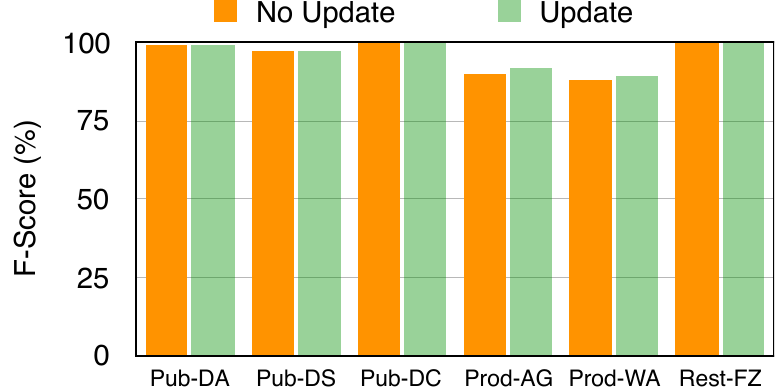}
      \vspace*{-4ex}
      \caption{Varying Vector Updates}
      \label{fig:varyingUpdate}
    \end{minipage}    
 \end{figure*}
 
Performing ER on a large dataset requires a number of design choices from the expert 
such as feature engineering, selection of appropriate similarity functions and thresholds, parameter tuning for ML models,
selection of appropriate blocking functions, and so on. 
Hence, it is incredibly hard to take any of the existing approaches and apply it as-is on a new dataset.
A key advantage of \sys is the ability to dramatically reduce this effort.
In order to highlight this feature, we evaluated \sys against Magellan~\cite{konda2016magellan} that also has a end-to-end EM pipeline.
We would like to emphasize that both \sys and Magellan share the dream of making the EM process as frictionless as possible.
While Magellan uses a series of sophisticated heuristics internally, \sys leverages \DRs as a foundational technique.
It is very easy to incorporate features from \sys and Magellan to each other.
For example, one can augment Magellan's automatically derived similarity based features to \sys while Magellan can readily use the blocking of \sys and so on. 
Table~\ref{tbl:deepERBestPerf} compares the performance of \sys and Magellan using their default settings.
Specifically, we adapted the end-to-end EM workflow for Magellan \cite{magellanWF}.
We can see that \sys beats Magellan on two datasets, while performing slightly worst in one datasets. Both systems delivered perfect results in the rather simple Fodors-Zagat dataset.

\stitle{Evaluating \sys for Other Domains.}
In order to show that our approach can be readily applied to other domains, 
we consider the problem of determining duplicates in a nucleotide database.
We assumed that we were provided with an appropriate dictionary for biomedical embeddings.
We evaluated our model on a large benchmark dataset~\cite{chen2016benchmarks} consisting of 21 
most heavily studied organisms in molecular biology.
Our method was able to beat ML models with hand-crafted approach in 11 of these cases 
while it was within a F1-score of $\pm$ 5 for the remaining.
Overall, our approach achieved an F1-measure of 87.4 for the automatic curation benchmark
where the state-of-the-art is 83.9.
This shows that our approach can be readily applied to other domains given the availability of effective embeddings.

  \begin{figure}[!t]
      \centering
      \includegraphics[scale=0.45]{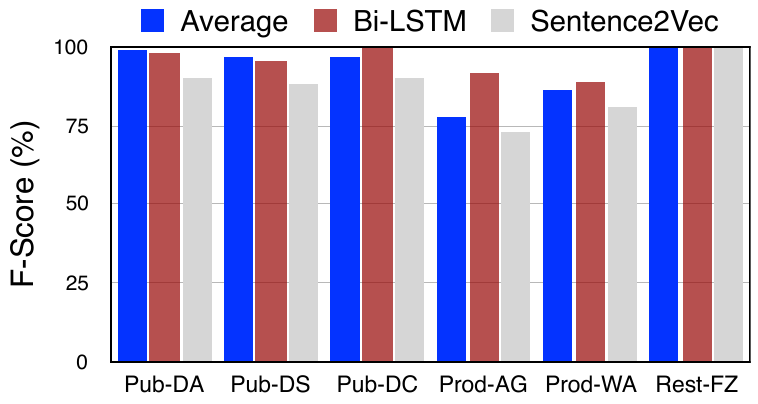}
      \vspace*{-3ex}
      \caption{Varying Composition}
      \vspace*{-3ex}
            \label{fig:varyingComposition}
 \end{figure}

\subsection{Understanding DeepER Performance}
\label{subsec:miscExpts}

We next investigate how each of the enhancements to the basic \sys architecture impacts its performance.
For this subsection, we use a positive to negative ratio of 1:4. 
This is different from the ratio we used when compared to competing approaches as this turned to be best for \sys.

\stitle{Varying the Size of Training Data.}
Figure~\ref{fig:varyingTrainingData} shows the results of varying the amount of training data.
\sys is robust enough to be competitive with other approaches with as little as 10\% of training data. 
Note that 10\% translates to as little as 11 examples 
to label in the case of Rest-FZ and to 543 in the case of Pub-DS. 
As expected, our method improves its excellent results with larger training data.
This is due to two key aspects: 
(a) the ability of \sys to leverage/transfer the semantic similarity learned by \DRs for ER and 
(b) a DL architecture that is customized for ER.
Please refer to~\cite{bengio2013representation,Goodfellow-et-al-2016,yosinski2014transferable} for additional details
about the beneficial impact of effective representation/transfer learning and well designed DL architectures.

\stitle{Impact of Incorrect Labels.}
Most of the prior work on ER assumes that the training data is perfect.
However, this assumption might not always hold in practice.
Given the increasing popularity of crowdsourcing for obtaining training data, 
it is likely that some of the labels for matching and non-matching pairs are incorrect.
We investigate the impact of incorrect labels in this experiment.
For a fixed set of training data (10\%), we vary the fraction of labels that are marked incorrectly.
Figure~\ref{fig:varyingNoise} shows the results.
While the F-measure reduces with larger fraction of incorrect labels, the experiments also show that our approach is very robust. 
The average drop in F-measure values compared to the perfect labeling case across all datasets at 10\% noise is just 2.6 with a standard deviation of 2.6. 
At 30\% the average drop is 8\% with a standard deviation of 7\%.
{We can also see that at 10\% noising, our approach is still competitive with state-of-the-art approaches. }

\stitle{Dynamic vs Static Word Embeddings.}
In this set of experiments, we evaluated the effect of updating (or fine-tuning) the initial word embeddings obtained from GloVe as part of training the model.
In other words, we evaluated if tuning the \DR for ER tasks improves the performance of our model.
Figure~\ref{fig:varyingUpdate} shows the results. 
The results matched our intuition that for the ``challenging'' datasets, updating the word embeddings in an end-to-end learning framework helped boost the results a little, 
whilst for the ``easy'' ones, it had either a small negative effect or no effect at all. 
Thus, we advise that in general, it is better to use the end-to-end framework.

\stitle{Varying Composition.}
In this set of experiments, we vary the compositional method we use to combine the individual word embeddings into a single representative vector for the tuple/attribute. 
In addition to word averaging and LSTM, we added a method based on Sentence2Vec~\cite{le2014distributed} and that is tuned using our end-to-end learning methods (Section~\ref{subsec:tunewe}), for completeness.
Figure~\ref{fig:varyingComposition} shows that for the ``easy'' datasets, simple word averaging work usually better than recurrent compositional models (LSTM or BiLSTM). 
This flips for the ``challenging'' datasets where sophisticated compositional approaches perform slightly better.
This is especially noticeable for  Prod-AG. 
Understanding and automatically recommending the appropriate architecture for a given dataset is a key focus of our future research.
In order to use the more complex compositional methods (LSTM or BiLSTM) one has to pay the price of its longer training times, one also has to tune its additional hyperparameters.
However, even the simple averaging compositional technique is competitive with prior approaches on all datasets.
We also observe that both methods are superior to Sentence2Vec with a slight exception on the Prod-AG dataset where it is superior to Average.

\stitle{Varying Word Embedding Dictionaries.}
Here, we study the impact of the dictionary used for \DRs.
GloVe has two major dictionaries : one trained on Common Crawl web corpus (840B tokens, 2.2M words) and one on Wikipedia (6B tokens, 400K words).
We used the vocabulary retrofitting to handle words not present in the dictionary.
Table~\ref{tbl:deepERVaryGlove} shows the result of the experiments.
As expected, there is a steep drop in F1 score when trained on a smaller dictionary.
The larger the corpus used for training the word embeddings, 
the better they are identifying semantic relationships.
\begin{table}[h!]
\vspace*{-3ex}
    \tabcaption{Impact of Word Embedding Dictionaries}
    \label{tbl:deepERVaryGlove}    
    {\small
    \begin{center}
        \begin{tabular}{|c|c|c|}
            \hline
            {\bf Dataset} & {\bf GloVe} & {\bf GloVe-Wiki}\\ \hline
            Pub-DA & 98.6 & 82.1 \\ \hline
            Pub-DS & 97.67& 77.8 \\ \hline
            Pub-DC & 99.1 & 79.2 \\ \hline
            Prod-WA & 88.06& 77.4 \\ \hline
            Prod-AG & 96.029& 87.2 \\ \hline
            Rest-FZ  & 100 & 91.2 \\ \hline
        \end{tabular}
    \end{center}
    }
      \vspace*{-4ex}    
\end{table}

\stitle{Varying Word Embedding Models.}
{We conducted experiments on three popular models, GloVe, Word2Vec, and FastText, 
which were trained
on a corpus with 840B tokens, 100B tokens and 600B tokens, respectively.
The number of words identified are 2.2M, 3M, and 1M, respectively.
Note that for fairness of comparison, we only considered word embeddings in FastText even though it also allows character embeddings.
We used the vocabulary retrofitting to handle words not present in the dictionary.
Table~\ref{tbl:deepERVaryGloveW2VecFastText} shows the results of the experiments.
In general, there are only minor variations between the different approaches.} 

\begin{table}[h!]
    \caption{{Impact of Word Embedding Used}}
    \label{tbl:deepERVaryGloveW2VecFastText}
    {
    \begin{center}
        \begin{tabular}{|c|c|c|c|}
            \hline
            {\bf Dataset} & {\bf GloVe} & {\bf Word2Vec} & {\bf FastText} \\ \hline
            Pub-DA & 98.6 & 97.9 & 98.2 \\ \hline
            Pub-DS & 97.6&  96.9 & 97.2 \\ \hline
            Pub-DC & 99.1 &  99 & 99\\ \hline
            Prod-WA & 88.06&  86.1 & 88.89\\ \hline
            Prod-AG & 96.03&  95.1 & 95.7\\ \hline
            Rest-FZ  & 100 &  100 & 100\\ \hline
        \end{tabular}
    \end{center}
    }
	\vspace*{-3ex}
\end{table} 

\stitle{Multi-Lingual Datasets.}
We now show an auxiliary benefit of using a \DR-based approach.
{We took three datasets that were originally in English and translated them to Spanish.}
We then used the \DRs for Spanish and repeated our approach.
Table~\ref{tbl:deepERVaryLanguage} shows that while there is a reduction in F1-score, 
our approaches can seamlessly work on multi-lingual datasets.
{Since we used ``Google translate'', we had to limit how much we could translate.
However, we expect to see similar results with 
the other datasets.
}
\begin{table}[h!]
      \vspace*{-2ex}    
    \tabcaption{{Showcasing ER on Multilingual Datasets}}
    \label{tbl:deepERVaryLanguage}
    {\small
    \begin{center}
        \begin{tabular}{|c|c|c|}
            \hline
            {\bf Dataset} & {\bf English} & {\bf Spanish}\\ \hline
            Prod-AG & 96.029& 89.1 \\ \hline
            Rest-FZ  & 100 & 92.6 \\ \hline
			{Pub-DS} & {97.67} & {88.1} \\ \hline
        \end{tabular}
    \end{center}
    }
      \vspace*{-3ex}    
\end{table} 
  

\subsection{Evaluating LSH-based Blocking}
\label{subsec:blockingExpt}

 \begin{figure*}[t!p]
         \centering
         \mbox{
                 \subfigure[K vs PC (L=10)]{\includegraphics[width=125pt]{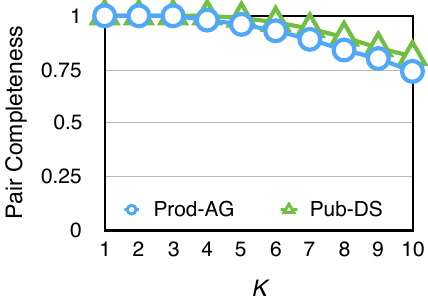}\label{subfig:blocking_vary_K_PC}}
                 \subfigure[K vs RR (L=10)]{\includegraphics[width=125pt]{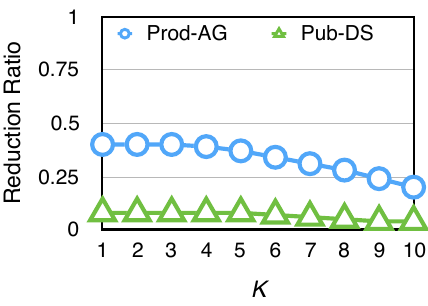}\label{subfig:blocking_vary_K_RR}}
                 \subfigure[L vs PC (K=4)]{\includegraphics[width=125pt]{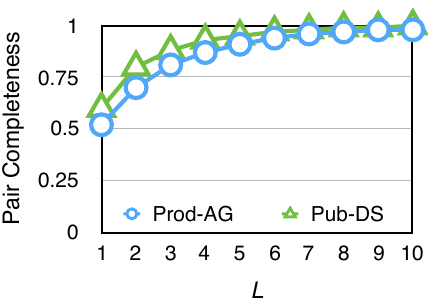}\label{subfig:blocking_vary_L_PC}}
                 \subfigure[L vs RR (K=4)]{\includegraphics[width=125pt]{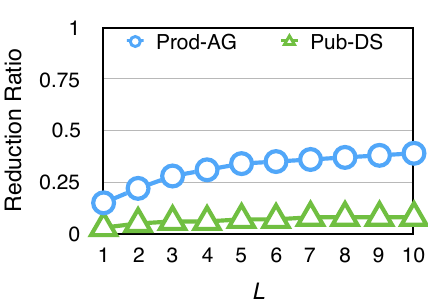}\label{subfig:blocking_vary_L_RR}}
         }
      \vspace*{-4ex}         
         \caption{Impact of varying K and L on Pair Completeness (PC) and Reduction Ratio (RR). 
         }
         \label{fig:blockingFigs}
      \vspace*{-2ex}         
 \end{figure*}

In this subsection, we evaluate the performance of our LSH based blocking approach.
Our approach allows us to vary $K$ (the size of the hash code) and $L$ (the number of hash tables) in order to achieve a tunable performance.
Recall that one can use Equation~\ref{eq:lshK} to derive $K$ and $L$ based on the task requirements.
Suppose we wish that similar tuples should fall into same bucket with probability $P_1=0.95$ and 
dissimilar tuples should fall into the same bucket with probability $P_2 \leq 0.5$.
Suppose that we index the DBLP dataset of Pub-DS.
Then based on Equation~\ref{eq:lshK}, we need a LSH with $K=12$ and $L=2$.

In our first set of experiments, we verify that the behavior of blocking is synchronous with the theoretical expectations.
We evaluate the performance of blocking based on two metrics widely used in prior research \cite{blocking,christen2012survey,michelson2006learning}.
The first metric, efficiency or reduction ratio (RR), is the ratio of the number of tuple pairs compared by our approach to the number of all possible pairs in $T$.
In other words, a smaller value indicates a higher reduction in the number of comparisons made. 
The second metric, recall or pair completeness (PC), is the ratio of the number of true duplicates compared by our approach against the total number of duplicates in $T$.
A higher value for PC means that our approach places the duplicate tuple pairs in the same block.

Figures~\ref{subfig:blocking_vary_K_PC}-\ref{subfig:blocking_vary_L_RR} show the results of our experiments.
As $K$ is increased, the value of PC decreases.
This is due to the fact that for a fixed $L$, increasing $K$ reduces the likelihood that two similar tuples will be placed in the same block 
which in turn reduces the number of duplicates that falls into the same block.
However, for a fixed $L$, increasing $K$ dramatically decreases the RR.
This is to be expected as a larger value of $K$ increases the number of LSH buckets into which tuples can be assigned to. 

A complementary behavior can be observed when we fix $K$ and vary $L$.
When $L$ increases,  PC also increases.
This is to be expected as the probability that two similar tuples being assigned to the same bucket increases when more than one hash table is involved.
In other words, even if a true duplicate does not fall into the same bucket in one hash table, it can fall into the same bucket in other hash tables.
However, increasing $L$ has a negative impact on RR as some false positive tuple pairs can fall into the same bucket in at least one hash table
thereby increasing the value of RR.

\begin{figure*}[t!p]
     \centering
     \mbox{
             \subfigure[K vs Recall (L=10)]{\includegraphics[width=125pt]{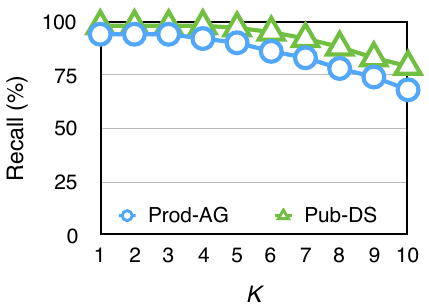}}
             \subfigure[K vs Precision (L=10)]{\includegraphics[width=125pt]{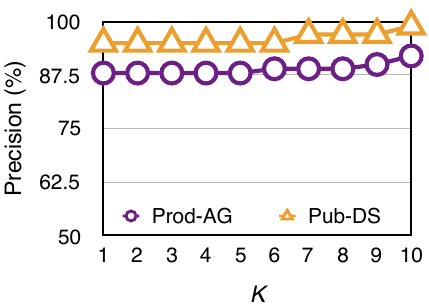}}
             \subfigure[L vs Recall (K=4)]{\includegraphics[width=125pt]{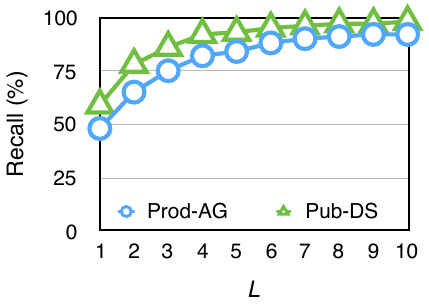}}
             \subfigure[L vs Precision (K=4)]{\includegraphics[width=125pt]{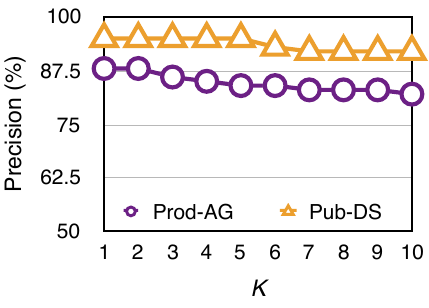}}
     }
     \caption{Impact of varying K and L on Precision and Recall of \sys }
     \label{fig:blockingEndToEnd}
\end{figure*}

\stitle{Evaluating \sys End-to-End.}
We next evaluate the performance of \sys by combining both the blocker and the matcher.
The results are shown in Figure~\ref{fig:blockingEndToEnd}.
First, we study how precision and recall are impacted by varying $K$ for a fixed $L$.
The recall decreases with increased $K$ as more and more duplicates are not put in the same block which results in \sys missing them.
The precision increases mildly with increasing $K$ as an increasing number of spurious non-duplicates are no longer being compared.
For example, when $K=1$, almost half of all possible pairs are classified by \sys, which reduces the precision with potential false positives.
However, when $K=10$, only a quarter of all possible pairs are classified, resulting in mild increase of precision.
The reason for the mild increase is that the classifier of \sys is relatively robust and achieved high precision even for low value of $K$.

Figures~\ref{fig:blockingFigs} c-d study the impact of varying $L$ for a fixed $K$ on precision and recall. 
As expected, the recall increases with higher $L$ as almost all the true duplicates are put in the same block and end up being classified as such by \sys.
Since \sys is quite accurate, this results in increased recall.
The precision declines mildly with increasing $L$.
The reason is that more and more non-duplicates are put in the same block resulting in potential false positives.
Once again, the impact is mild as the classifier is relatively robust.

\begin{figure}[h!]
	\centering
	\includegraphics[scale=0.66]{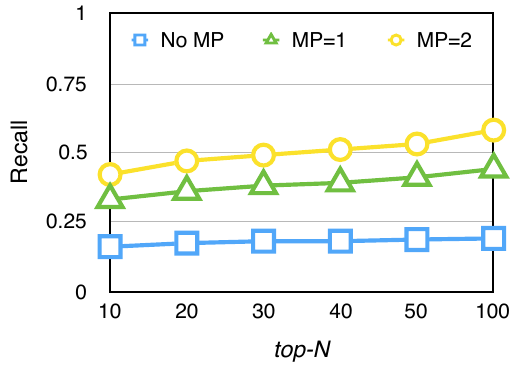}
    \caption{MultiProbe LSH on Pub-AG}
	\label{fig:topKMP}
\end{figure}

\stitle{Evaluating Multi-Probe LSH.}
We evaluate Algorithm~\ref{alg:lshANNSearch} using Multi-probe and comparing a tuple only with top-$N$ most similar tuples instead of all tuples in a block.
Figure~\ref{fig:topKMP} shows the result for Pub-AG.
We vary the number of multi-probes and pick the top-N most similar tuples to be classified.
We measure the recall of this approach for $K=10$ and an extreme case with a {\em single} hash table where $L=1$.
We wish to highlight two trends.
First, even using a single multi-probe sequence can dramatically increases the recall.
This supports our claim that one can increase recall using a small number of hash tables by using multi-probe LSH.
Second, increasing the size of $N$ does not dramatically increase the recall.
This is due to the fact that duplicate tuples have high similarity between their corresponding distributed representations.
Our top-$N$ based approach would be preferable to reduce the number of classifier invocations when the block size is much higher than 10.

%

\section{Related Work}
\label{sec:related}

\stitle{Entity Resolution.}
A good overview of ER can be found in surveys such as~\cite{DBLP:journals/tkde/ElmagarmidIV07,naumann2010introduction}.
Prior work can be categorized as based 
on (a) declarative rules, (b) ML and (c) expert or crowd based.
Declarative rules, such as DNF which specify rules for matching tuples, are easily interpretable~\cite{synthesizer} but often requires a domain expert.
Most of the ML approaches are variants of the classical Fellegi-Sunter model~\cite{theoryrecordlinkage}.
Popular approaches include SVM~\cite{DBLP:conf/kdd/BilenkoM03}, active learning~\cite{SarawagiB02},
clustering~\cite{DBLP:conf/kdd/CohenR02}.
Recently, ER using crowdsourcing has become popular~\cite{DBLP:journals/pvldb/WangKFF12,corleone}.
While there exist some work for learning similarity functions and thresholds~\cite{DBLP:conf/kdd/BilenkoM03,DBLP:journals/pvldb/WangLYF11},
ER often requires substantial involvement of the expert.

There has been extensive work on building EM systems.
\cite{konda2016magellan} provides a comprehensive survey of the current EM systems.
Most of the prior works often do not cover the entire EM pipeline, require extensive interaction with experts and are not turn-key systems.
The key objective of \sys is the same as Magellan~\cite{konda2016magellan}. 
We propose a end-to-end EM system based on \DR that minimizes the burden on the experts.
Our techniques are modular enough and can be easily incorporated into any of the existing systems.

{
There has been extensive interest in applying DL in data cleaning~\cite{thirumuruganathan2018data}. 
A recent work that extends \sys~\cite{anhaisigmod2018} explored the design space of ER using \DRs.
The authors introduce four 
different choices for the attribute summarization process,
namely, SIF, RNN, Attention, and Hybrid.
The first two methods are similar to our AVG and LSTM-RNN methods.
Attention uses  decomposable attention for attribute summarization and 
vector concatenation  to perform attribute comparison.
Hybrid uses a bidirectional RNN with decomposable attention for attribute summarization and a 
vector concatenation and element-wise absolute difference 
during attribute comparison. 
While this work shares some similarities with \sys, there are two notable differences.
We present solutions for the practical situations where 
for dealing with data with partial or minimal coverage.
We also propose efficient and effective blocking solutions. 

\stitle{Blocking.}
Blocking has been extensively studied as a way to scale ER systems and 
a good overview can be found in surveys such as \cite{blocking, christen2012survey}.
Common approaches include key-based blocking that partitions tuples into blocks based on their values on certain attributes and 
rule-based blocking where a decision rule determines which block a tuple falls into.
There has been limited work on simplifying this process by 
either learning blocking schemes such as \cite{michelson2006learning} or tuning the blocking \cite{kenig2013mfiblocks}.
In contrast, our work automates the blocking process by requiring minimal input from the domain expert.

Some recent works used LSH for blocking.
\cite{steorts2014comparison} uses MinHashing where tuples with high Jaccard similarity 
are likely to be assigned to the same block. 
\cite{wang2016semantic} improves it by proposing a MinHashing with semantic similarity based on concept hierarchy to assign conceptually similar tuples to the same block.
\cite{fisher2015clustering} proposed a clustering based method to satisfy size constraints
with upper and lower size thresholds for blocks for performance and privacy reasons.


\section{Final Remarks}
\label{sec:conclusion}

In this paper, we introduced \sys, a DL-based approach for ER.
Our fundamental contribution is the identification of the concept of \DRs 
as a key building block for designing effective ER classifiers.
We also propose algorithms to transform a tuple to a \DR, 
building \DR-aware classifiers and an efficient blocking strategy based on LSH.
Our extensive experiments show that our approach is promising and already 
achieves or surpasses the state-of-the-art on multiple benchmark datasets.
We believe that DL is a powerful tool that has applications in databases beyond ER
and it is our hope that our ideas be extended to build practical and effective ER systems.

There are several avenues for improving \sys.
First, understand and automatically recommend the appropriate DL architecture for a given dataset, especially complex ones.
Another line of work is to  design a hybrid system that leverages both 
automatic features, such as \DRs,  and manual features, such as a similarity metric for IDs.
Finally, we will also need to address the cases where the data is dirty.

\balance
\interlinepenalty=10000
\bibliographystyle{abbrv}
\bibliography{DA}

\end{document}